\documentclass[letterpaper]{article}
\usepackage[T1]{fontenc}

\usepackage{hyperref}
\usepackage[backend=biber,style=chem-acs,sorting=none,articletitle=true,chaptertitle=true,doi=true,url=true,eprint=true]{biblatex}
\usepackage{xurl}
\usepackage{xpatch}
\DeclareFieldFormat[inproceedings]{title}{#1}
\xpatchbibdriver{inproceedings}
  {\usebibmacro{in:}}
  {\usebibmacro{title}\setunit{\addperiod\space}\usebibmacro{in:}}
  {}{\PackageError{consolv-bibliography}{Cannot enable conference-paper titles}{Check the chem-acs bibliography driver.}}
\usepackage{xr-hyper}

\makeatletter
\newcommand*{\addFileDependency}[1]{%
  \typeout{(#1)}%
  \@addtofilelist{#1}%
  \IfFileExists{#1}{}{\typeout{No file #1.}}%
}
\makeatother
\addFileDependency{supplementary_information_acs.aux}

\usepackage{geometry}
\usepackage{setspace}

\usepackage{amsmath}
\usepackage{amssymb}
\usepackage{mathrsfs}
\usepackage{subcaption}
\usepackage{caption}
\usepackage{tabularx}
\usepackage{booktabs}
\usepackage{xcolor}
\definecolor{green}{RGB}{0,120,0}
\usepackage{multirow}
\usepackage{diagbox}
\usepackage{makecell}
\usepackage{adjustbox}
\usepackage{pifont}
\usepackage{ragged2e}
\usepackage{enumerate}
\usepackage{todonotes}
\usepackage{algpseudocode}
\usepackage{algorithm}
\usepackage{graphicx}
\usepackage{float}
\newfloat{scheme}{htbp}{los}
\floatname{scheme}{Scheme}
\graphicspath{{./}}

\usepackage{authblk}
\author[1]{Linying Zhang}
\author[1,2]{Julija Zavadlav\textsuperscript{*}}
\affil[1]{Multiscale Modeling of Fluid Materials, Department of Engineering Physics and Computation, TUM School of Engineering and Design, Technical University of Munich, 85748 Garching, Germany}
\affil[2]{Atomistic Modeling Center (AMC), Munich Data Science Institute (MDSI), Technical University of Munich, 85748 Garching, Germany}
\date{*Email: julija.zavadlav@tum.de}

\title{ConSolv: Solvent-Conditional Machine Learning Implicit Solvent Potential}

\begin{document}

\maketitle



\begin{abstract}
Implicit solvent machine learning potentials (MLPs) offer a powerful route to bridging the gap between accuracy and efficiency in molecular simulations. However, existing models have largely focused on aqueous environments, overlooking the diverse and important roles of non-aqueous solvents in areas such as organic synthesis and battery technology. Here, we present ConSolv, a solvent-conditional MLP architecture that explicitly incorporates solvent effects on solute interactions through an attention-based solvent-embedding block. By combining experimental solvation free energy data with ab initio data, we train a single implicit solvent MLP that is transferable across 66 common organic solvents. ConSolv outperforms classical explicit solvent methods and selected ab initio implicit solvent approaches across multiple solvation free energy benchmarks, and demonstrates generalization to unseen solvents. Beyond solvation free energies, the model shows close agreement with experimental nuclear magnetic resonance (NMR) data for $\gamma$-fluorohydrin molecules in chloroform. ConSolv's architecture is readily extensible to broader chemical spaces and alternative training strategies, while its attention-based design supports explainable artificial intelligence (AI) analysis that can help elucidate complex, solvent-dependent molecular interactions.
\end{abstract}

\section{Introduction}\label{sec:intro}
Solvents play an active and often decisive role in molecular processes. They modulate intermolecular interactions, screen electrostatic forces, stabilize or destabilize intermediates and transition states, and shift reaction equilibria. These effects are central to a wide range of applications, including pharmacology, organic synthesis, and electrochemistry~\cite{hirata2003molecular, gorges2022towards, matczak2017heteroatom, odey2023unraveling, ensing2001solvation}, where the solvent environment influences drug solubility, reaction rates, electrolyte stability, and the conformational dynamics of solute molecules. Accurately capturing both solvent and solute dynamics is therefore essential for improving the predictive power of molecular dynamics (MD) simulations.

Conventional simulation methods, however, face a fundamental trade-off between accuracy and efficiency. Quantum-mechanical approaches such as density functional theory (DFT) can provide high accuracy, but their computational cost is often prohibitive to study solvent-mediated molecular processes. Classical force fields such as GAFF~\cite{vassetti2019assessment}, by contrast, enable explicit-solvent simulations over larger spatial and temporal scales, but rely on fixed functional forms and approximate parameterizations that may limit their accuracy and transferability. Machine learning (ML) potentials (MLPs)~\cite{zuo2020, mueller2020, Friederich2021Machine}, particularly those based on graph neural networks (GNNs), have emerged as a powerful route to overcoming this trade-off. By learning flexible many-body interaction potentials directly from data, MLPs can approach ab initio accuracy at a fraction of the computational cost. Recent advances have therefore enabled explicit-solvent simulations with near-ab initio accuracy~\cite{ple2025foundation}, opening new opportunities for the atomistic study of solvent effects.

Despite these advances, explicit-solvent MLP simulations remain computationally demanding. Although MLPs scale linearly with the number of particles and can therefore be applied to very large systems, including systems containing millions of atoms~\cite{stormer2026aluminum}, they remain orders of magnitude slower than classical force fields. This makes it challenging to reach the long timescales needed to sample solvent-driven conformational dynamics, reaction pathways, and rare events. Implicit-solvent MLPs offer a promising solution to this bottleneck. By removing explicit solvent degrees of freedom and encoding solvent effects directly into the effective solute interactions, they can provide speedups of two to three orders of magnitude, particularly when solvent molecules constitute the majority of the system~\cite{Rocken:24HFE}. Contrary to traditional models~\cite{zhou2002, chen2021, cumberworth2016free}, ML-based implicit-solvent models can, in principle, reproduce explicit-solvent conformational ensembles of complex molecular and macromolecular systems with high fidelity, provided that sufficient training data and adequate model expressivity are available~\cite{thaler2022deep}. Nonetheless, developing transferable implicit solvent MLPs requires addressing two intertwined questions: how the solvent effects should be incorporated into the architecture, and which data should be used to parametrize the model.

On the architectural level, one strategy is to use ML only for the solvent-solute contribution, while treating intramolecular interactions with conventional methods. Within this framework, parameterization can be physics-based or fully data-driven. Physics-based approaches leverage established implicit solvent models such as Generalized Born~\cite{wojciechowski2004generalized} and use ML to optimize their parameters~\cite{mahmoud2020generalized, katzberger2023implicit, katzberger2025transferring}. For example, the GNNIS model uses a GNN to predict environment-dependent Generalized Born radii and solvent-dependent surface-tension coefficients for 39 organic solvents~\cite{katzberger2025rapid}. Such models benefit from physical grounding and are less likely to produce unphysical behavior outside the training distribution. The data-driven approaches~\cite{chen2021, airas2023transferable} instead learn effective solvent-solute interactions directly from simulation data, offering greater flexibility but typically requiring broader training sets or unconventional training approaches~\cite{thaler2022deep,ding2022contrastive,durumeric2026learning}. While both parameterization strategies benefit from reduced system complexity, they share a fundamental limitation, as their accuracy is bounded by the fidelity of the traditional method used to model the solute. The alternative strategy overcomes this limitation by allowing the MLP to model the intramolecular interactions, which are modified to account for the solvent environment. This approach offers the greatest flexibility and, in principle, can achieve the highest accuracy, but existing models of this type have so far focused exclusively on aqueous environments~\cite{Rocken:24HFE, mutlu2025data, andris2026machine, ding2022contrastive, charron2025navigating}, leaving the chemically diverse and technologically important space of non-aqueous solvents largely unexplored.

The choice of training data is equally critical. The first option is to train on explicit-solvent simulations, which requires coarse-graining over the solvent degrees of freedom. Unlike conventional MLPs, where configurations can be generated with an inexpensive MD simulation and subsequently labeled with a high-fidelity (e.g., DFT) model, coarse-grained models must be trained on configurations generated by the target model itself, since configurations sampled under a different potential introduce a statistical bias unless the configurational ensemble is reweighted~\cite{chen2025enhanced}. Since training directly on ab initio explicit-solvent data is computationally prohibitive, previous approaches relied on classical force fields as data sources~\cite{katzberger2025rapid}, which inherently limits the attainable accuracy. A second option is to train on implicit solvent data. This avoids coarse-graining but transfers the approximations of the underlying implicit solvent model to the MLP. Semi-empirical quantum implicit-solvent methods, including COSMO-RS~\cite{mu2008conductor} and self-consistent continuum solvation (SCCS) approaches~\cite{Hille2019}, that treat the solute quantum mechanically and the solvent as a dielectric continuum could be employed. However, large-scale implicit-solvent datasets comparable in size and chemical diversity to SPICE~\cite{eastman2023spice} or OMol25~\cite{levine2025open} are not yet available, and existing datasets such as Aquamarine~\cite{sandonas2024aquamarine} are largely restricted to water. A third option is to train on experimental observables. This top-down strategy circumvents the need for expensive quantum-mechanical solvent calculations, but requires differentiable links between model parameters and experimental quantities. For example, the DiffTRe framework~\cite{Thaler:21} enables this link via reweighting for time-independent observables. The recent ReSolv model~\cite{Rocken:24HFE} extended DiffTRe to parameterize an implicit water MLP by combining DFT data for gas-phase molecules with experimental hydration free energies. This approach achieved hydration-free-energy predictions more accurate than those of classical explicit-solvent models while providing speedups of four orders of magnitude relative to explicit-solvent MLPs. Nevertheless, the extensions to non-aqueous solvents are still missing.

In this work, we present ConSolv, a solvent-conditional implicit-solvent MLP that is transferable across small organic molecules. Instead of developing a separate model for each solvent, ConSolv models multiple non-aqueous solvents within a single, unified architecture, enabling information transfer between solvents. The attention-based solvent-embedding block is used to incorporate the solvent effects by modulating solute interactions in a solvent-dependent manner while maintaining the flexibility of a fully data-driven MLP. Building on the ReSolv parametrization, we integrate DFT data for molecules in vacuum with experimental solvation free energy data from Solv@TUM~\cite{Hille2019}, allowing for simultaneous training across 66 non-aqueous solvents. We demonstrate that ConSolv accurately predicts solvation free energies for unseen solute molecules across multiple benchmark datasets, generalizes to new solvents, and offers interpretable solvent-solute interaction patterns through its attention mechanism. By extending implicit-solvent MLPs beyond aqueous environments, ConSolv establishes a scalable and flexible framework for investigating solvent-dependent molecular behavior across common solvents.

\section*{Methods}\label{sec:method}

\subsection*{Message Passing Graph Neural Network Potential}
Message Passing Graph Neural Networks (MPGNNs)~\cite{gilmer2017neural} are state-of-the-art architectures for predicting potential energy of a system based on the positions and chemical elements of atoms. Any MPGNN architectures could be extended to include solvent conditioning. Here, we use the MACE model~\cite{batatia2022mace}, a popular equivariant MPGNN, as a baseline architecture. Below, we summarize the main components, while detailed information can be found in Refs.~\cite{batatia2022mace,kovacs2025mace}.

\begin{figure}[h!]
    \centering
    \includegraphics[width=1.0\textwidth]{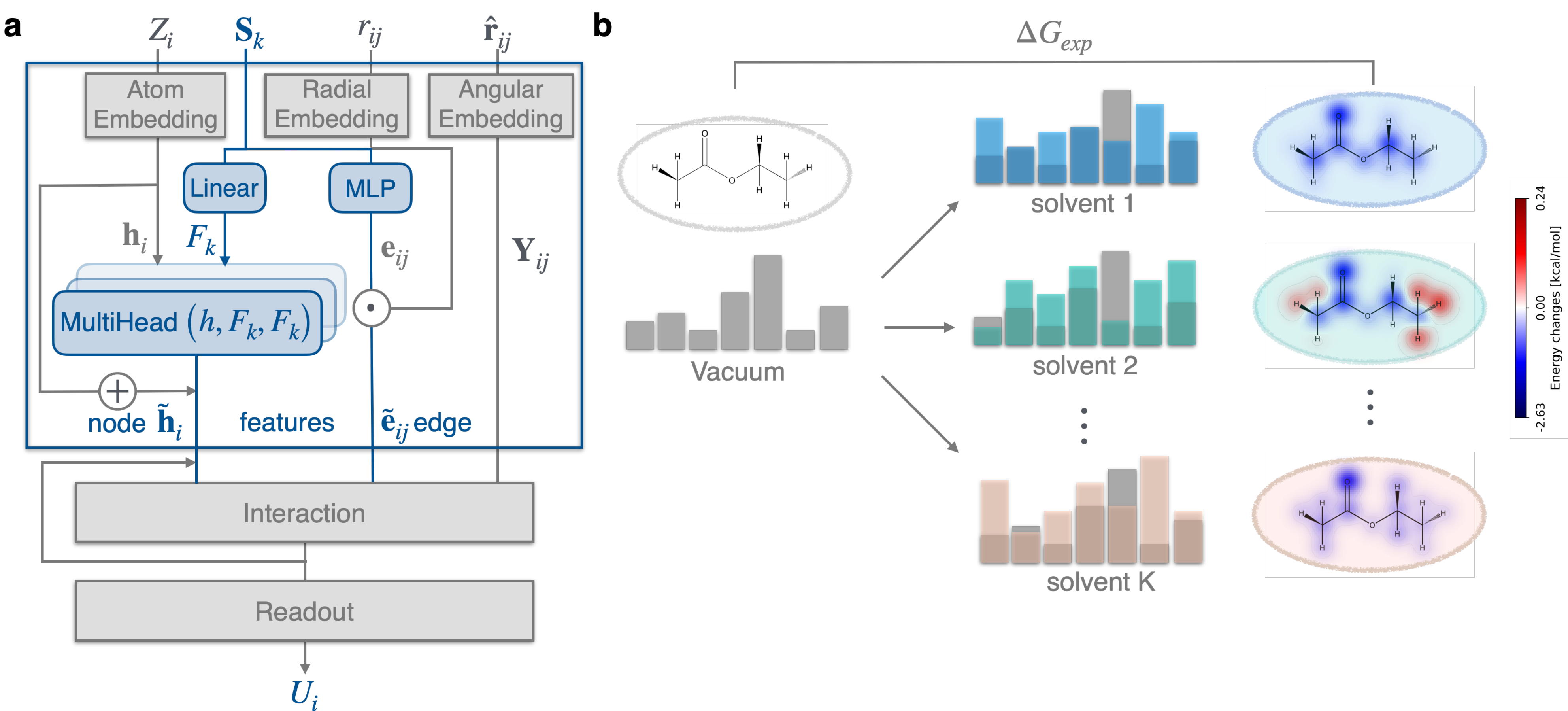}
    \caption{ConSolv architecture (a) and training (b). In a conventional message passing graph neural network (MPGNN; a, gray), the node features $\mathbf{h}_i$ initially represent atom types $Z_i$. The atomic environment is encoded with radial $\mathbf{e}_{ij}$ and angular $\mathbf{Y}_{ij}$ embeddings, where $r_{ij}$ and $\hat{\mathbf{r}}_{ij}$ denote the interatomic distance and unit direction vector, respectively. The solvent embedding block (a, blue) outputs the modulated node $\tilde{\mathbf{h}}_i$ and edge $\tilde{\mathbf{e}}_{ij}$ features represented with bar plots (b) based on the $k$-th solvent descriptor vector $\mathbf{S}_k$ via multi-head attention and multi-layer perceptron (MLP), respectively. Consequently, the per-atom energy $U_i$ changes depending on the solvent, as visualized for the ethyl propanoate molecule in (top to bottom) dimethylformamide, chloroform, and octanol for the RDKit-generated minimum energy conformer. The parameters are trained to match the experimental solvation free energy $\Delta G_{exp}$.
    }
    \label{fig:concept}
\end{figure}

MPGNNs operate on a graph, in which each atom defines a node and an edge is drawn between atoms $i$ and $j$ when their separation $r_{ij}$ is smaller than a predefined cutoff radius $r_{\text{cutoff}}$, defining the local environment of each atom. The node features $\mathbf{h}_i\in\mathbb{R}^{N_{\text{channel}}}$ are initialized with an atom embedding block as trainable vectors that encode atom types $Z_i$ into $N_{\text{channel}}$ channels, obtained via index embedding based on the atom's position in the periodic table (Figure~\ref{fig:concept}a).
In the radial embedding block, edge features $\mathbf{e}_{ij}(r_{ij})$ are generated from interatomic distances, with each $r_{ij}$ expanded in a Bessel radial basis and multiplied by a smooth polynomial cutoff envelope before being passed to a small multilayer perceptron, yielding distance-dependent embeddings that decay smoothly to zero at the cutoff. The angular embedding $\mathbf{Y}_{ij}(\hat{\mathbf{r}}_{ij})$ encodes directional information via spherical harmonics.

The interaction block iteratively updates node features via message passing, aggregating information from neighboring nodes and edges. At each message passing layer $t$, a message is constructed for each atom
\begin{equation}
\mathbf m_i^{(t)}=\sum_{j\in\mathcal N(i)} \phi_m^{(t)}\left(\mathbf h_i^{(t)},\mathbf h_j^{(t)},\mathbf{e}_{ij}, \mathbf{Y}_{ij} \right),
\end{equation}
where $\phi_m^{(t)}$ is a learnable message function. The permutation-invariance is ensured by the pooling operation over neighboring atoms $j\in\mathcal N(i)$, while tensor products are used to ensure the resulting representations transform equivariantly under rotation. The node features are updated using
\begin{equation}
    \mathbf{h}_i^{(t+1)} = \phi_u^{(t)}\left(\mathbf{h}_i^{(t)},  \mathbf{m}_{i}^{(t)}\right),
\end{equation}
where $\phi_u^{(t)}$ is a learnable update function. After $T$ message passing iterations, a readout function aggregates the node features to predict per-atom energy contribution
\begin{equation}
U_i=\sum_{t=1}^T \phi_r^{(t)}\left( \mathbf{h}_i^{(t)}\right),
\end{equation}
where $\phi_u^{(t)}$ is a learnable readout function. The forces on the atoms are calculated by taking analytical derivatives of the total potential energy with respect to the positions of the atoms.

\subsection*{ConSolv architecture}
 ConSolv extends the MPGNN architecture to enable a solvent-conditional potential energy model. In particular, we encode the $k$-th solvent with a feature vector $\mathbf{S}_k \in \mathbb{R}^{N_d}$, where subscript $k=1,\dots,N_s$ runs over solvent type. Classical descriptors based on solvent molecular structure and properties are used to construct the solvent embedding feature vector $\mathbf{S}_k$. The selection of molecular descriptors was guided by the findings of Ferraz-Caetano \textit{et al}.~\cite{ferraz2023explainable}, who demonstrated that the accurate prediction of solvation Gibbs free energy is primarily driven by the solvent's and solute's polar surface area, polarizability, and electronic charge distribution. Consequently, we employed $N_d=15$ standard RDKit~\cite{rdkit}-generated 2D and 3D descriptors, including Van der Waals Surface Area descriptors, molecular refractivity, and atomic partial charges, that effectively capture fundamental physicochemical properties of molecules. The descriptor set is detailed in {Supporting Information Table~S2}.

 We use precomputed solvent features to modulate the MPGNN embedding block such that the node and radial features depend not only on atom type $Z_i$ and radial distance $r_{ij}$ but also on the solvent descriptor $\mathbf{S}_k$ (Fig.~\ref{fig:concept}a). The node feature modulation is implemented using a multi-head cross-attention mechanism~\cite{vaswani2017attention}. The attention mechanism computes a weighted sum of value vectors, where the weights are determined by the compatibility between a query and a set of keys
\begin{equation}
    \text{Attention}(Q, K, V) = \text{softmax}\left(\frac{QK^\top}{\sqrt{d}}\right)V,
    \label{eq:attention}
\end{equation}
where $Q$, $K$, and $V$ denote the query, key, and value matrices, respectively. $d$ is the dimension of the queries and keys used to stabilize the dot product magnitude. Each solvent descriptor vector $\mathbf{S}_k$ is projected to a $p$-dimensional space using a linear layer, resulting in a solvent embedding matrix $F_k \in \mathbb{R}^{N_d \times p}$. The atom embedding $\mathbf{h}_i$ serves as the query. In contrast, the solvent embedding matrix $F_k$ serves as both the key and value, allowing each atom to attend selectively to the most relevant solvent features. To capture diverse solvent-solute interactions simultaneously, $H$ parallel attention heads are used. The modulated atom embedding $\tilde{\mathbf{h}}_i$ is obtained by adding the attention-weighted solvent embedding to the original atom embedding $\mathbf{h}_i$, i.e.,
\begin{align}
    \tilde{\mathbf{h}}_i &= \mathbf{h}_i + \text{Concat}(\text{head}_1, \ldots, \text{head}_m, \dots, \text{head}_H)W^O, \\
    \text{head}_m &= \text{Attention}(\mathbf{h}_iW^Q_m, F_k W^K_m, F_k W^V_m),
    \label{eq:multihead}
\end{align}
where $W^Q_m \in \mathbb{R}^{N_{\text{channel}} \times d}$, $W^K_m \in \mathbb{R}^{p \times d}$, $W^V_m \in \mathbb{R}^{p \times d_v}$, and $W^O \in \mathbb{R}^{H d_v \times N_{\text{channel}}}$ are learnable matrices, with $d_v$ denoting a dimension of the value matrix.

For the edge features $\mathbf{e}_{ij}$, a multi-layer perceptron $g(\cdot)$ is employed to transform the solvent descriptors $\mathbf{S}_k$ to produce a modulation factor for each component of the Bessel radial basis embedding
\begin{equation}
    \tilde{\mathbf{e}}_{ij} = \mathbf{e}_{ij} + g(\mathbf{S}_k) \odot \mathbf{e}_{ij},
\end{equation}
where $\odot$ denotes element-wise multiplication.

The modulated node features $\tilde{\mathbf{h}}_i$ and edge features $\tilde{\mathbf{e}}_{ij}$ are the outputs of the solvent embedding block. The node features $\tilde{\mathbf{h}}_i$ are passed to the first interaction (message-passing) layer, and the edge features are passed to all interaction layers. The spherical harmonic edge attributes $\mathbf{Y}(\hat{\mathbf{r}}_{ij})$ remain unchanged.

\subsection*{Two Stage Training using ReSolv}
We train the ConSolv MLP architecture using the ReSolv method previously employed to obtain an implicit water MLP model~\cite{Rocken:24HFE}. Briefly, ReSolv consists of two stages: a bottom-up and top-down training. In the first stage, we start from a randomly initialized parameter set and perform a standard training against a DFT dataset using the energy and force matching loss function
\begin{equation}
    L_{1} = \frac{1}{N_{bs}} \sum^{N_{bs}}_{l=1} \left[ w_U (U_l-U_{l, \text{DFT}})^2 +
    w_F \frac{1}{3N_{l,a}} \sum^{N_{l,a}}_{i=1} \sum^{3}_{m=1}(F_{ilm}-F_{ilm, \text{DFT}})^2\right],
\end{equation}
where $w_U$ and $w_F$ are user-defined weights for the energy $U_l$ and force $F_{ilm}$ on atom $i$ in configuration $l$ and direction $m$, respectively. The subscript DFT denotes the ab initio data target, $N_{bs}$ batch size, while $N_{l,a}$ is the number of atoms in configuration $l$. After training, we obtain an MLP model that takes a configurational and chemical state $\mathcal{C}$ as input and predicts the potential energy $U_{vac}=U(\mathcal{C};\theta_{vac})$ for molecules in vacuum with near-ab initio accuracy.

In the second stage, the architecture is augmented with a solvent-embedding block. The resulting MLP model is solvent-type-dependent, i.e., $U_{solv}=U(\mathcal{C}, S;\theta_{solv})$, where $S$ is the solvent type. The parameter set is correspondingly enlarged $\theta_{solv} = \{\theta, \theta_{se}\}$, where $\theta$ parameters are initially set to the values of the vacuum model, while the parameters of the solvent embedding block $\theta_{se}$ are initialized randomly using a Gaussian distribution with zero mean and a standard deviation of 0.1. This initialization introduces a small perturbation to the vacuum model and enables accurate free energy calculations from trajectories sampled from the vacuum state. After initialization, all parameters of the model $\theta_{solv}$ are optimized using a loss function that minimizes the difference between the calculated free energy difference $\Delta G$ between the vacuum model $U(\mathcal{C};\theta_{vac})$ and the implicit solvent model $U_{solv}(\mathcal{C}, S;\theta_{solv})$ and the experimental solvation free energy $\Delta G_{exp}$
\begin{equation}
    L_{2}=\frac{1}{N_{bs}}\sum_{l=1}^{N_{bs}}\left(\Delta G-\Delta G_{exp}\right)^2.
\end{equation}
This training process continues until the model converges, yielding the final implicit solvent potential $U_{solv}$.

Second-stage training is not as straightforward as the first stage, because $\Delta G$ is not a direct output of an MLP but is instead computed via MD simulations. Nevertheless, we can use the fact that free energy is a state function and construct a free energy path along the training procedure, i.e., $\Delta G=\Delta G_{\theta_{vac} \to \theta_{solv}} = \Delta G_{\theta_{vac} \to \tilde{\theta}} + \Delta G_{\tilde{\theta} \to \theta_{solv}}$, where $\tilde{\theta}$ corresponds to parameters of the implicit solvent model at some point during training and $\theta_{solv}$ are the current parameters. The first term has been computed during previous training steps, while the second term can be estimated from a reference trajectory $\{\mathcal{C}_n\}_{n=1}^N$ generated under reference parameters $\tilde{\theta}$ using
\begin{equation}
    \Delta G_{\tilde{\theta} \to \theta_{solv}} = -\beta^{-1} \ln\left(N^{-1}\sum_{n=1}^N \exp\left\{-\beta\left[U_{solv}(\mathcal{C}_n,S;\,\theta_{solv}) - U_{solv}(\mathcal{C}_n,S;\,\tilde{\theta})\right]\right\}\right),
\end{equation}
where $\beta = 1/k_B T$. This provides a differentiable relation between $\Delta G$ and $\theta_{solv}$, enabling gradient-based optimization. To reduce the cost of trajectory generation, reference trajectories are reused across training steps via an effective sample size criterion~\cite{carmichael2012new}. When the criterion is no longer satisfied, a new trajectory is generated, and the accumulated free energy difference is updated using the Bennett Acceptance Ratio (BAR) method~\cite{shirts2003equilibrium}. For further details, see Ref.~\cite{Rocken:24HFE}.

\subsection*{Datasets and Further Details}
We use a filtered version of the DFT-generated SPICE dataset~\cite{moore_kovacs_browning_batatia_horton_kapil_witt_magdau_cole_csanyi_2024} for the first stage. SPICE contains over $10^{5}$ distinct molecular systems covering drug-like molecules and peptides, and over $2\times10^{6}$ conformations. Charged and ill-posed molecules are excluded, resulting in a dataset covering 10 chemical elements \{H, C, N, O, F, P, S, Cl, Br, I\}, corresponding to 85\% of the original dataset size~\cite {eastman2023spice}. The selected data is randomly split into train (95\%) and test (5\%) portions. After training, the resulting MLP for molecules in vacuum achieves test-set accuracies of 1.8 meV/atom and 46 meV/Å for energies and forces, respectively. The accuracy is below the chemical accuracy threshold of 1 kcal/mol (43.4 meV) and comparable to that of state-of-the-art models trained on the same dataset. For example, the MACE-OFF23(S)~\cite{kovacs2025mace} achieves errors of 1.0--2.0 meV/atom and 16--40 meV/\AA~on a matching test set. Slightly lower force errors of the MACE-OFF23(S) model compared to ours could be due to a larger number 
of trainable parameters.

For the second stage, we use the Solv@TUM database~\cite{Hille2019}, containing 5952 experimental solvation free energy values for neutral molecules across 146 non-aqueous solvents. For consistency with the vacuum MLP, molecules containing elements outside the filtered SPICE dataset are excluded, yielding 5592 unique solute-solvent pairs. We select a subset of 66 solvents with more than 30 solute data points to avoid overfitting, resulting in 4577 data points. ConSolv is initially trained on all of these points to check the overall error distribution. We observe systematic outliers for molecules with positive solvation free energies (Fig.~S1). These outliers correspond to small gas molecules: deuterium, nitrogen monoxide, carbon monoxide, oxygen, carbon tetrafluoride, and nitrogen. We attribute this behavior to two reasons: (i) the SPICE dataset contains drug-like organic molecules and peptides, and thus, the vacuum MLP cannot extrapolate to gas molecules; (ii) the experimental solvation free energies of soluble gases are difficult to measure accurately~\cite{wilhelm1977low}. Since our model is intended for organic and drug-like molecules, gas molecules were subsequently excluded from the dataset. Additionally, molecules for which the vacuum model yields unstable simulations, identified by the presence of isolated atoms beyond the cutoff distance of 5.0~\AA ~during trajectory generation, are also excluded. After all filters, solute molecules within each solvent are randomly split into 80\% for training and 20\% for testing, resulting in 3596 training and 923 test data points.

The MD simulations performed during training are initiated with solute configurations generated by OpenBabel~\cite{o2011open} from the Simplified Molecular Input Line Entry System (SMILES) strings in the Solv@TUM database. We perform a 25 ps initial equilibrium followed by a 225 ps production run for each molecule. We use the velocity Verlet algorithm with a 1~fs time step. Simulations are performed in the constant-number, constant-volume, constant-temperature (NVT) ensemble with a Langevin thermostat at 298.15~K and a friction coefficient of 50 ps$^{-1}$. The same protocol is used for training and testing. The convergence of the solvation free energy estimate using these settings was previously tested~\cite {Rocken:24HFE}. All MD simulations are conducted using Jax MD~\cite{schoenholz2020jax}. The MLP training was performed using chemtrain software~\cite{fuchs2025chemtrain}. All hyperparameter settings regarding the  MLP architecture and numerical optimization are reported in Supporting Information Table~S1.

\section*{Results} \label{sec:results}

\subsection*{ConSolv outperforms explicit solvent force fields in predicting Solvation Free Energies}
We first test ConSolv's performance in predicting solvation free energy on different held-out data (Table~\ref{tab:com_sccs}). To enable extensive comparison with traditional models, including DFT-based implicit solvent models SCCS~\cite{Hille2019} and COSMO-RS~\cite{klamt1998refinement} as well as the explicit solvent classical force field GAFF~\cite{zhang2015force}, we use three datasets: the Solv@TUM test set,
CombiSolv~\cite{vermeire2021transfer}, and Zhang \textit{et al}.~\cite{zhang2015force} benchmark. Across all datasets, the estimated experimental uncertainties are around 0.2~kcal/mol~\cite{vermeire2021transfer,zhang2015force}, providing a reference noise floor for interpreting the reported errors.
\begin{table}[t]

\captionsetup{skip=8pt}
\centering
\footnotesize
\setlength{\tabcolsep}{3pt}
\renewcommand{\arraystretch}{1.25}
\caption{Solvation free energy prediction accuracy of ConSolv against reference methods. RMSE and MAE are reported in kcal/mol; bold indicates the best value within each test set. \emph{Matched} indicates whether every method was evaluated on the same solute--solvent pairs; \emph{Overlap with training set} records whether the test pairs appear in the ConSolv training set. Provenance: $^{a}$computed in this work; $^{b}$literature aggregate; $^{c}$metrics recomputed from published predictions. n.r.\ = not reported; n.d.\ = not determinable.}
\label{tab:com_sccs}
\begin{tabularx}{\linewidth}{@{}>{\RaggedRight\arraybackslash}p{0.13\linewidth}lrrrrc>{\Centering\arraybackslash}p{0.12\linewidth}>{\RaggedRight\arraybackslash}X@{}}
\toprule
Test set & Method & $N$ & RMSE & MAE & $R^{2}$ & Matched
  & \makecell{Overlap with\\training set} & Subset definition \\
\midrule
\multirow[t]{2}{=}{Solv@TUM~\cite{Hille2019}}
  & ConSolv$^{a}$  & 923  & {\textbf{0.48}} & {\textbf{0.31}} & 0.97 & \ding{55} & None
  & \multirow[t]{2}{=}{Held-out split defined separately per method} \\
  & SCCS$^{b}$     & n.r. & 0.84~\cite{Hille2019} & n.r. & n.r. & \ding{55} & n.d. & \\
\addlinespace[8pt]
\multirow[t]{2}{=}{CombiSolv~\cite{vermeire2021transfer}}
  & ConSolv$^{a}$  & 471 & {0.80} & {0.57} & {0.94} & \checkmark & None
  & \multirow[t]{2}{=}{$\leq 11$ heavy atoms; no gas or iodine solutes} \\
  & COSMO-RS$^{c}$ & 471 & \textbf{0.71} & \textbf{0.40} & \textbf{0.95} & \checkmark & None & \\
\addlinespace[8pt]
\multirow[t]{3}{=}{Zhang \textit{et al.}~\cite{zhang2015force}}
  & ConSolv$^{a}$  & 50 & {\textbf{0.69}} & {0.53} & \textbf{0.93} & \checkmark & None
  & \multirow[t]{3}{=}{50 of 228 published pairs} \\
  & GAFF$^{c}$     & 50 & 1.18 & 0.87 & 0.85 & \checkmark & None & \\
  & COSMO-RS$^{c}$ & 50 & 0.71 & \textbf{0.50} & 0.92 & \checkmark & None & \\
\bottomrule
\end{tabularx}
\end{table}

\begin{figure}[h!]
    \centering
    \includegraphics[width=1.0\linewidth]{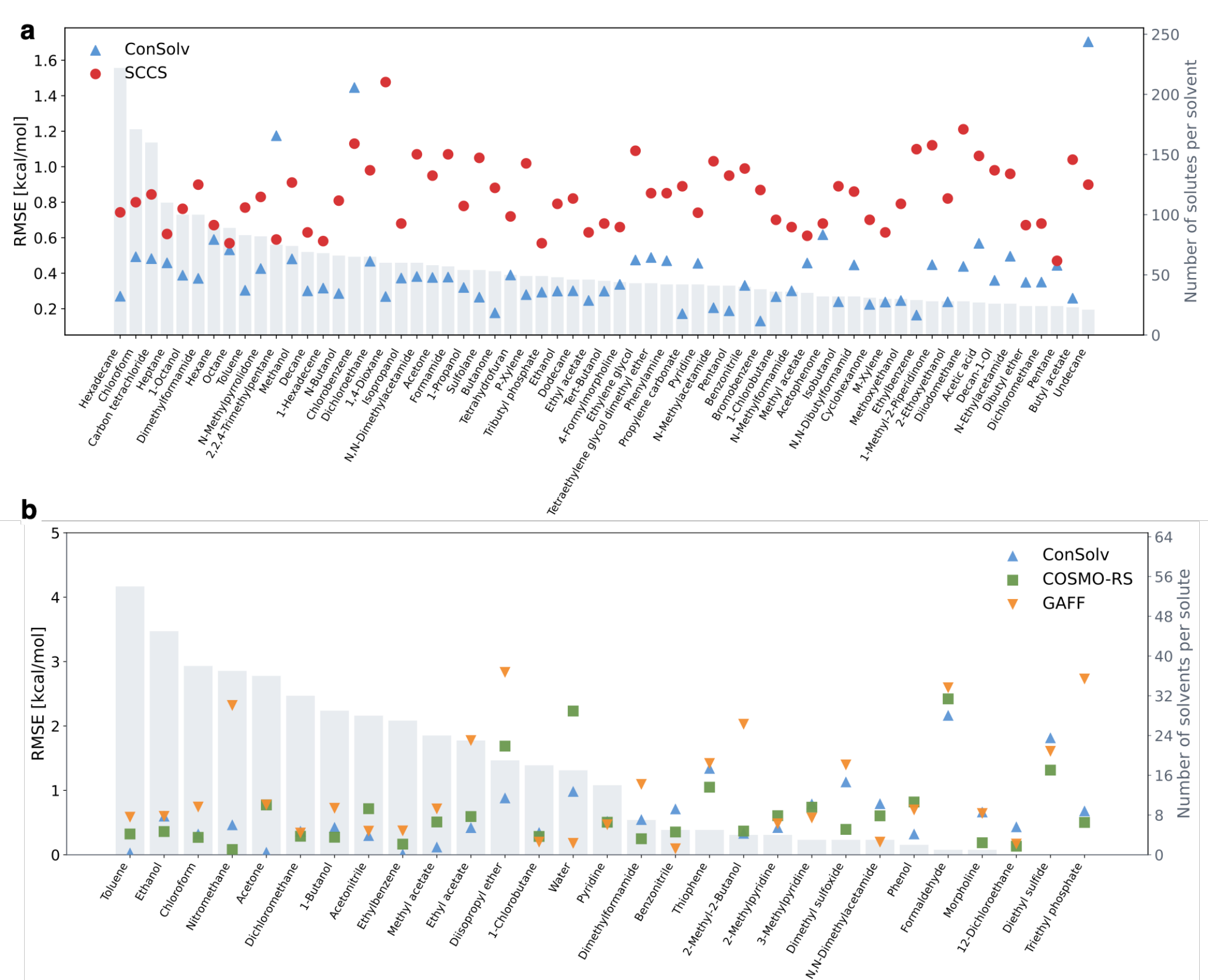}
    \caption{Root mean squared error (RMSE) with respect to solvent using Solv@TUM test set (a) and solute using the Zhang {\textit{et al.}}~\cite{zhang2015force} benchmark (b). The corresponding training dataset size is shown as a gray bar plot. Results are shown for ConSolv (blue), SCCS (red)~\cite{Hille2019}, COSMO-RS (green)~\cite{zhang2015force}, and GAFF (orange)~\cite{zhang2015force} models.}
    \label{fig:accuracy}
\end{figure}

On the Solv@TUM test set (923 points), ConSolv reliably predicts solvation free energies, {with over 95\% of ConSolv's predictions within 1~kcal/mol of experimental values}. Moreover, {ConSolv has a lower reported RMSE than SCCS on separately defined held-out splits} (Table~\ref{tab:com_sccs}).
On the Solv@TUM {training set} (3596 points), the MAE is 0.31 kcal/mol, the RMSE is 0.45 kcal/mol, and the $R^2$ is 0.97. We identified two outliers with errors above 4~kcal/mol, visible in the parity plot (Fig.~S3a), namely the iodine molecule in chlorobenzene and in 2,2,4-trimethylpentane solvents. These errors stem from the low representation of iodine-containing molecules in the SPICE training set, resulting in an inaccurate vacuum potential energy surface and, consequently, large errors in solvation free energy. These two outliers directly cause the RMSE spikes for chlorobenzene and 2,2,4-trimethylpentane in the per-solvent breakdown ({Fig.~\ref{fig:accuracy}a}). Excluding the iodine molecule from the test set reduces the RMSE for both solvents below 1 kcal/mol.
Beyond these outliers, ConSolv achieves lower errors than SCCS for most solvents, with the only remaining solvent above 1~kcal/mol being undecane, which has limited representation during training with only 24 data points. To further assess the robustness of ConSolv beyond individual solvents, we analyze the error distributions across different chemical spaces and physicochemical properties. {Partitioning the accuracy metrics by solvent’s functional group reveals no meaningful error divergence despite unequal data availability, confirming broad applicability across the organic chemical space (Fig.~S5a). The predictive performance remains consistent across solvent polarity classifications, with a uniform error distribution across polar protic, polar aprotic, and non-polar environments (Fig.~S5c).} Examination by {solute's functional group} shows high accuracy for common chemical groups and slightly larger variance for specialized ones, consistent with the data distributions in the training set (Fig.~S5b). Finally, categorization by solute's molecular size reveals a gradual increase in MAE with the number of heavy atoms, reflecting the higher complexity of larger molecules and their relative scarcity in the training data (Fig.~S5d).

Next, we evaluate ConSolv against COSMO-RS using a subset of the CombiSolv database~\cite{vermeire2021transfer}. To assess extrapolation beyond the Solv@TUM training distribution, we retain solute--solvent pairs with both experimental and COSMO-RS reference values that are absent from the Solv@TUM training set. For consistency with the chemical space targeted during ConSolv training, we further exclude solutes with more than 11 heavy atoms, gas molecules, and iodine-containing molecules, which were identified as outliers in the preceding evaluations. The final filtered test set contains 471 data points. COSMO-RS achieves superior accuracy owing to its quantum mechanical treatment of the solute, which requires a DFT-level calculation per molecule and is therefore computationally demanding (Table~\ref{tab:com_sccs}). Nevertheless, ConSolv retains good predictive accuracy, with an overall MAE below the chemical-accuracy threshold. ConSolv also captures the overall trend in solvation free energies for this external CombiSolv benchmark subset (Fig.~S3b), although the error distribution indicates a slight systematic overestimation of solvation free energies (Fig.~S3e). {For datapoints with experimental solvation free energies below -10 kcal/mol, which correspond to larger solute molecules, the simulation length is extended from 250 ps to 1 ns. The resulting solvation free energies differed between these two simulation lengths by only 0.004 kcal/mol, which eliminates the effect of insufficient sampling.} Results for a less restrictive filtered dataset, which includes solutes with up to 19 heavy atoms and contains 661 data points, are reported in the Supporting Information. On this set, ConSolv achieves an MAE of 0.67~kcal/mol and an RMSE of 0.95~kcal/mol. The larger errors mainly arise because ConSolv is primarily trained on molecules with fewer than 11 heavy atoms, leading to increased errors for larger solutes (Fig.~S6b), and because the experimental solvation free energies of larger molecules are also larger in magnitude (Fig.~S6a). When the averaged absolute relative error (AARE) is considered instead of the absolute error, the performance is more consistent across molecular sizes (Fig.~S6c). {The residual increase in absolute error with molecular size is consistent with the decreasing representation of larger and more flexible solutes in the SPICE and Solv@TUM training data~\cite{eastman2023spice,Hille2019}. Because both the vacuum intramolecular potential and the solvent-dependent correction are learned from these data, we restrict the validated domain of the current model to small neutral organic molecules represented in the training chemical space.}

Lastly, we compare ConSolv with GAFF and COSMO-RS using the benchmark proposed by Zhang \textit{et al.}~\cite{zhang2015force}. This benchmark contains 228 data points, of which 50 solute-solvent pairs are not included in the ConSolv training dataset, while the solvent is one of the 66 solvents for which ConSolv is parameterized. On this subset, ConSolv substantially outperforms GAFF, as expected, considering that GAFF was not parameterized to reproduce solvation free energies. Examining the per-solute predictive errors (Fig.~\ref{fig:accuracy}b) highlights divergent failure modes. For GAFF, molecules containing nitro functional groups, including nitromethane and nitrobenzene, show pronounced deviations, arising from the limitations of assigning static partial charges to highly polar constituents, which cannot adequately encode complex electrostatic interactions and polarization effects~\cite{zhang2015force}. For ConSolv, the largest predictive errors occur for molecules with high polarizability or highly constrained internal geometries, such as diethyl sulfide and thiophene. In addition, the RMSE for ConSolv {increases as the amount of available training data per solute decreases}, reflecting the typical data limitations of machine learning models. Despite these data limitations, ConSolv reaches an overall accuracy close to that of COSMO-RS, and for several solutes it yields lower predictive errors (Fig.~\ref{fig:accuracy}b). This agreement is notable because COSMO-RS evaluates each solute using a quantum-mechanical description, whereas ConSolv maintains comparable performance after learning from sparse solute-specific training data. The per-solvent error analysis (Fig.~S4) confirms that ConSolv predictions are reliable, with nearly all solvent-specific RMSE values close to or below 1~kcal/mol, while GAFF introduces large statistical variance and significant errors for several solvents.

\subsection*{Importance of the solvent embedding semantics}
The ConSolv architecture is not the only approach to making an MLP potential solvent-dependent. Many architectural variations are possible, and the optimal choice likely depends on the specific application and dataset. Prior studies have not systematically explored different solvent embeddings for MLPs, leaving the optimal architecture an open question. Designing a solvent-dependent MLP framework requires decisions on three key elements: the stage at which the embedding is applied, the embedding method, and the mechanism for modulating graph features. We compare five architectural variants across these three design axes. The embedding stage determines where solvent information is incorporated: \textit{input-based} approaches modulate node and edge features before message passing, while the \textit{output-based} approach applies modulation only to the final per-atom potential energy after the readout layer. The embedding method specifies how the solvent is represented: \textit{descriptor-based} methods use fixed, predetermined physicochemical solvent descriptors, whereas \textit{trainable} methods learn a distinct embedding vector for each solvent directly from data. The modulation mechanism dictates how the solvent representation is combined with solute graph features, and we examine both \textit{attention} and \textit{concatenation} mechanisms. ConSolv employs the input-based descriptor approach with attention modulation. The four alternative combinations evaluated are: input-based descriptor with concatenation, input-based trainable with attention, input-based trainable with concatenation, and output-based descriptor with attention. Implementation details for these variants are provided in the Supporting Information.

\begin{figure}[h!]
    \centering
    \includegraphics[width=1.\linewidth]{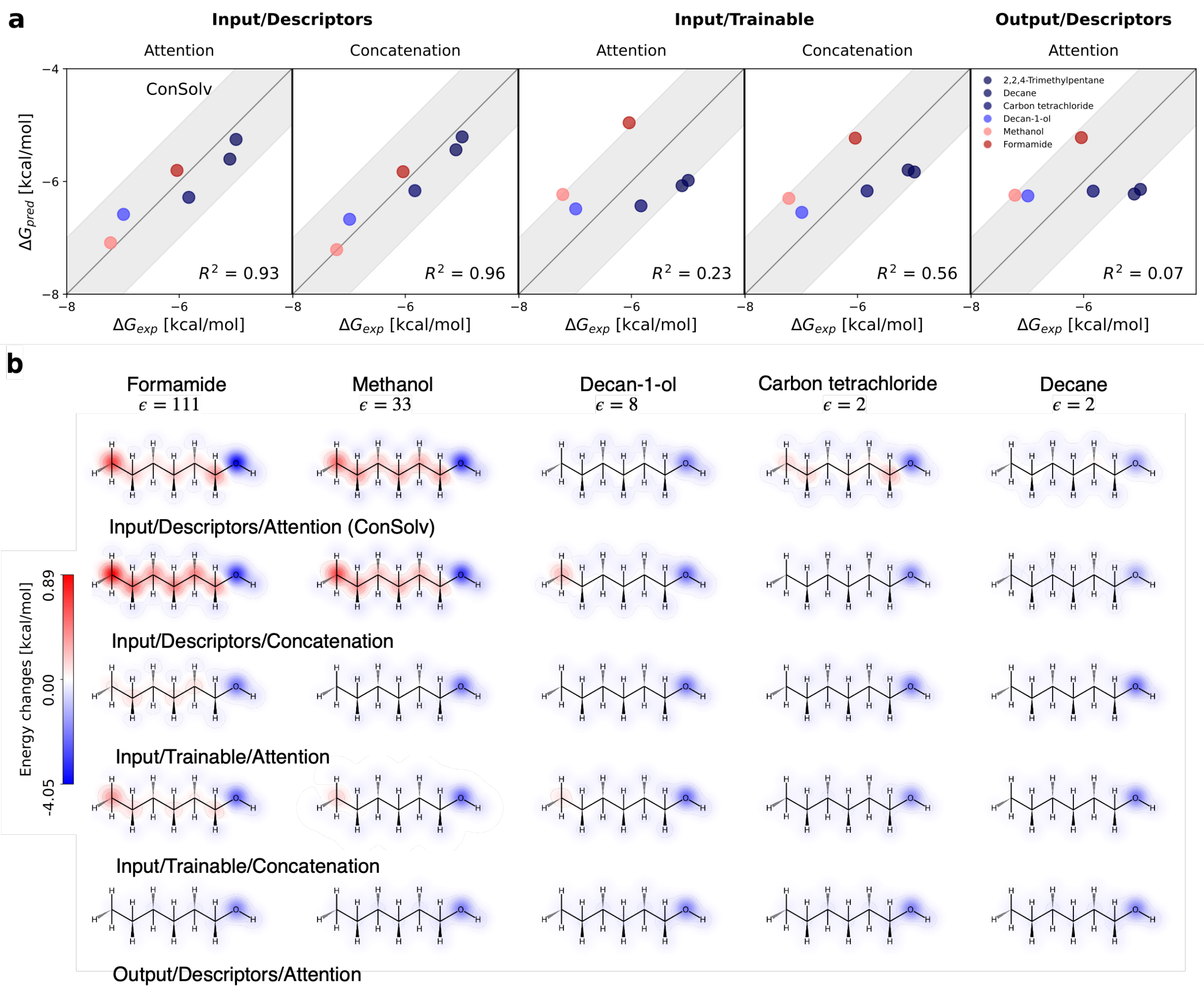}
    \caption{(a) Parity plot of solvation free energy predictions for 1-hexanol solute molecule across six solvents with diverse dielectric constant denoted by the color of the points with blue and red representing the lowest and highest dielectric constants, respectively. The black line represents the ideal prediction, while the shaded gray region marks the chemical accuracy region (1 kcal/mol). {(b) Trajectory-averaged per-atom potential energy contribution differences between the vacuum and solvent models.} The red and blue colors refer to positive and negative changes, respectively, while white refers to the zero value. {The results are shown for 1-hexanol in five solvents with the solvent dielectric constant decreasing from left to right.}}
    \label{fig:structure_comparison}
\end{figure}

{The input-based descriptor embedding with concatenation achieves the lowest RMSE, the lowest MAE, and the highest per-solute $R^2$ on the Solv@TUM test set (Table~\ref{tab:com_struc}). Both descriptor-based input models distinguish solvent environments more effectively than the trainable-embedding and output-based variants, as reflected by their higher per-solute $R^2$ values. ConSolv's attention-based design retains strong aggregate accuracy while providing a directly interpretable solvent-modulation mechanism. We therefore retain ConSolv as the preferred architecture because it balances predictive accuracy with physically meaningful modulation of interatomic interactions (Figure~\ref{fig:structure_comparison}b) and interpretability through the attention mechanism (Figure~\ref{fig:attn_analysis}).}
\begin{table}[h!]
    \centering
    \caption{Impact of solvent embedding architecture design on the Solv@TUM test set performance. Root mean squared error (RMSE) and mean absolute error (MAE) are in units of kcal/mol. The Pearson correlation coefficient $R^2$ per solute is calculated with the weighted Fisher's z-transformation~\cite{fisher1921probable}.}
    \begin{tabular}{r|r|r|c|c|c|c}
        Embedding & Embedding & Modulation & RMSE & MAE & $R^2$ & $R^2$ \\
        Stage & Method & Mechanism &  &  & per solute & overall \\
        \hline
        (ConSolv) Input & Descriptors & Attention
        & {0.48} & {0.31} & {0.84} & \textbf{0.97} \\ \hline
        Input & Descriptors & Concatenation
        & {\textbf{0.42}} & {\textbf{0.27}} & {\textbf{0.88}} & \textbf{0.97} \\
        Input & Trainable & Attention
        & {0.58} & {0.39} & {0.72} & 0.95 \\
        Input & Trainable & Concatenation
        & {0.55} & {0.37} & {0.73} & 0.95 \\
        Output & Descriptors & Attention
        & {0.70} & {0.51} & {0.35} & 0.92
    \end{tabular}
    \label{tab:com_struc}
\end{table}

To substantiate our findings, we analyze the models' predictive patterns for representative solutes in the test dataset. Figure~\ref{fig:structure_comparison}(a) displays the key advantage of the descriptor-based approach, namely its ability to resolve solvent-specific variations. For 1-hexanol solute molecule evaluated across six solvents, the descriptor-based models closely reproduce experimental trends, while both input-based trainable embedding models and the output-based model predict nearly identical solvation free energies across all solvents. This demonstrates a clear inability to distinguish between different solvent environments. Similar shortcomings are observed for other solutes in the test set (Figure~S2). These failures can be attributed to two factors: trainable embeddings rely solely on learned representations, making them prone to overfitting and poor generalization to unseen solutes, while output-based solvent embeddings are incorporated too late in the MPGNN to influence the message-passing encoding of local atomic environments.

Figure~\ref{fig:structure_comparison}b further illustrates these differences by visualizing the per-atom potential energy changes, $\Delta U = U_{solv} - U_{vac}$. {The shifts were evaluated over configurations sampled from the ConSolv production trajectory for each solvent. The reported value for atom $i$ was the trajectory average, $\langle\Delta U_i\rangle=N_{\mathrm{frame}}^{-1}\sum_{f=1}^{N_{\mathrm{frame}}}\Delta U_i^{(f)}$. The index $i=1,\ldots,N_{\mathrm{atom}}$ runs over all atoms of the solute, and $f=1,\ldots,N_{\mathrm{frame}}$ runs over the saved production frames for that solvent. Here, $N_{\mathrm{frame}}=500$. For each frame $f$, all learned models and the vacuum model were evaluated using the same atomic coordinates, and the per-atom solvent-induced shift was calculated as $\Delta U_i^{(f)}=U_{i,\mathrm{solv}}^{(f)}-U_{i,\mathrm{vac}}^{(f)}$.} Although per-atom potential energies are not rigorous quantum-mechanical observables, their solvent-induced variations provide qualitative insight into how each architecture redistributes atomic energy contributions in response to solvent information. ConSolv exhibits physically consistent and solvent-sensitive energy redistributions that agree with established group-contribution descriptions of solvation phenomena~\cite{cramer1999implicit}. As solvent polarity increases (Figure~\ref{fig:structure_comparison}b right to left), carbon atoms along the alkyl chain shift from weakly stabilizing negative shifts to small positive corrections. This trend is consistent with the increasing free-energy cost associated with cavity formation and reduced favorable dispersion interactions for nonpolar aliphatic groups in highly cohesive polar solvents~\cite{reichardt2011solvents,cramer1999implicit}. Simultaneously, the hydroxyl group exhibits increasing stabilization from low- to high-dielectric solvents due to increasingly favorable electrostatic interactions with the surrounding medium. {The fixed-coordinate GBn2 polar electrostatic analysis~\cite{nguyen2013gbn2}, evaluated on the same ConSolv configurations (Figure~S10), supports this interpretation by similarly localizing the strongest stabilization around the hydroxyl group, with a stronger response in higher-dielectric solvents and comparatively weaker contributions along the alkyl chain. Differences in magnitude and detailed spatial distribution are expected because GBn2 represents only the polar electrostatic contribution, whereas the learned energy shifts contain the full solvent-dependent model response.} {The Input/Descriptors/Concatenation architecture also exhibits a clear polarity-dependent redistribution, including increasingly positive contributions along the alkyl chain in formamide and methanol together with persistent hydroxyl stabilization. However, its response along the hydrocarbon chain is stronger and more spatially distributed than that observed for ConSolv. The Input/Trainable/Attention architecture shows a substantially weaker solvent dependence, with stabilization remaining primarily localized around the hydroxyl group and appreciable changes along the alkyl chain appearing mainly in formamide. The Input/Trainable/Concatenation architecture displays a similarly limited solvent response, with only modest redistribution of the alkyl-chain contributions across solvents. In contrast, the Output/Descriptors/Attention architecture produces nearly invariant energy fields across the solvent series, maintaining similar hydroxyl-centered stabilization with little solvent-dependent variation along the hydrocarbon chain.}

\subsection*{Generalization to unseen solvents}
\begin{figure}[h!]
    \centering
    \footnotesize
    \includegraphics[width=1.0\linewidth]{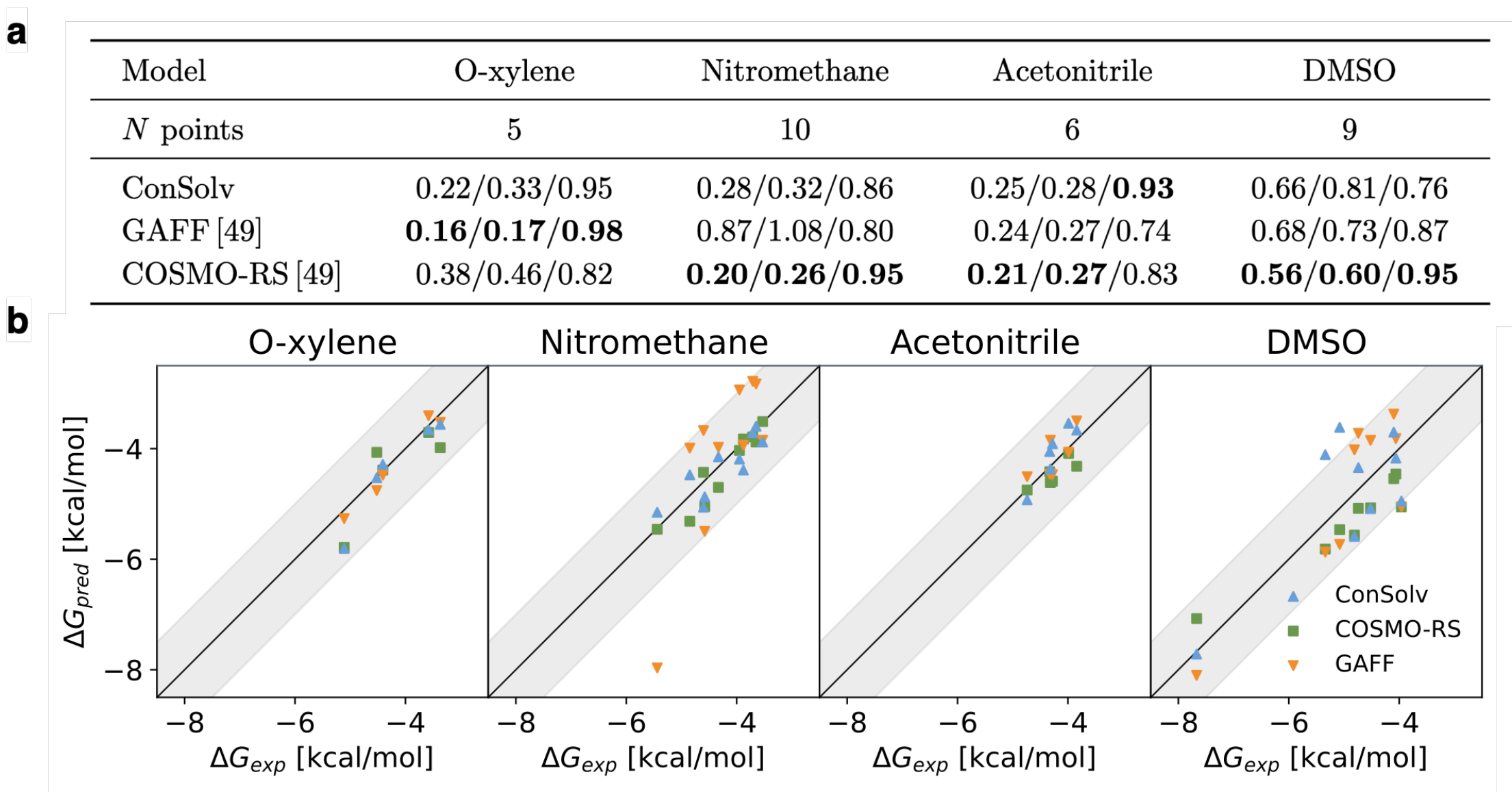}
    \caption{Generalization to solvents unseen during ConSolv training (o-xylene, acetonitrile, dimethyl sulfoxide (DMSO), nitromethane) evaluated using $N$ experimental data points~\cite{zhang2015force}. Error metrics (a) including root mean squared error (RMSE, kcal/mol), mean absolute error (MAE, kcal/mol), and Pearson correlation $R^2$, and parity plots (b) for ConSolv, GAFF~\cite{zhang2015force}, and COSMO-RS~\cite{zhang2015force} models.}
    \label{fig:unseen}
\end{figure}

A key advantage of descriptor-based solvent representations over trainable solvent embeddings is their capacity to predict properties for entirely new solvents without requiring retraining. To assess this generalization, we evaluate ConSolv on solvents absent from the training set, thus testing its transferability beyond the 66 organic solvents for which it was originally parametrized. Specifically, we investigate the performance on the four holdout solvents o-xylene, acetonitrile, dimethyl sulfoxide (DMSO), and nitromethane using, in total, 30 experimental data points extracted from Zhang \textit{et al.}~\cite{zhang2015force}. These solvents cover a wide spectrum of physicochemical properties, ranging from weakly polar aromatic environments (o-xylene) to highly polar aprotic solvents (DMSO and nitromethane). ConSolv exhibits strong predictive accuracy across all tested solvents (Fig.~\ref{fig:unseen}a).
Compared to GAFF and COSMO-RS, it achieves the highest $R^2$ for acetonitrile. Notably, ConSolv's predictions do not show significant outliers across all evaluated solvents (Fig.~\ref{fig:unseen}b). In contrast, the GAFF simulations exhibit a pronounced outlier in nitromethane solvent corresponding to the nitromethane's self-solvation free energy. Nitromethane possesses a strong permanent dipole together with highly localized charge separation around the nitro group. Since conventional fixed-charge force fields rely on static partial charges, they cannot explicitly adapt to environment-dependent polarization effects in strongly polar media~\cite{mobley2007comparison}. In pure nitromethane, strong dipole-dipole interactions can therefore become excessively stabilized if the assigned charges slightly overestimate the interaction strength, a known limitation for generalized force-field treatments of nitro-containing systems~\cite{zhang2015force,zhang2017}. Consequently, GAFF produces an artificially over-stabilized solvation free energy. Overall, the results indicate that ConSolv exhibits far greater reliability in highly polar solvents compared to GAFF.

\subsection*{Generalization to out-of-target experimental data}
\begin{figure}[h!]
    \centering
    \includegraphics[width=1.0\linewidth]{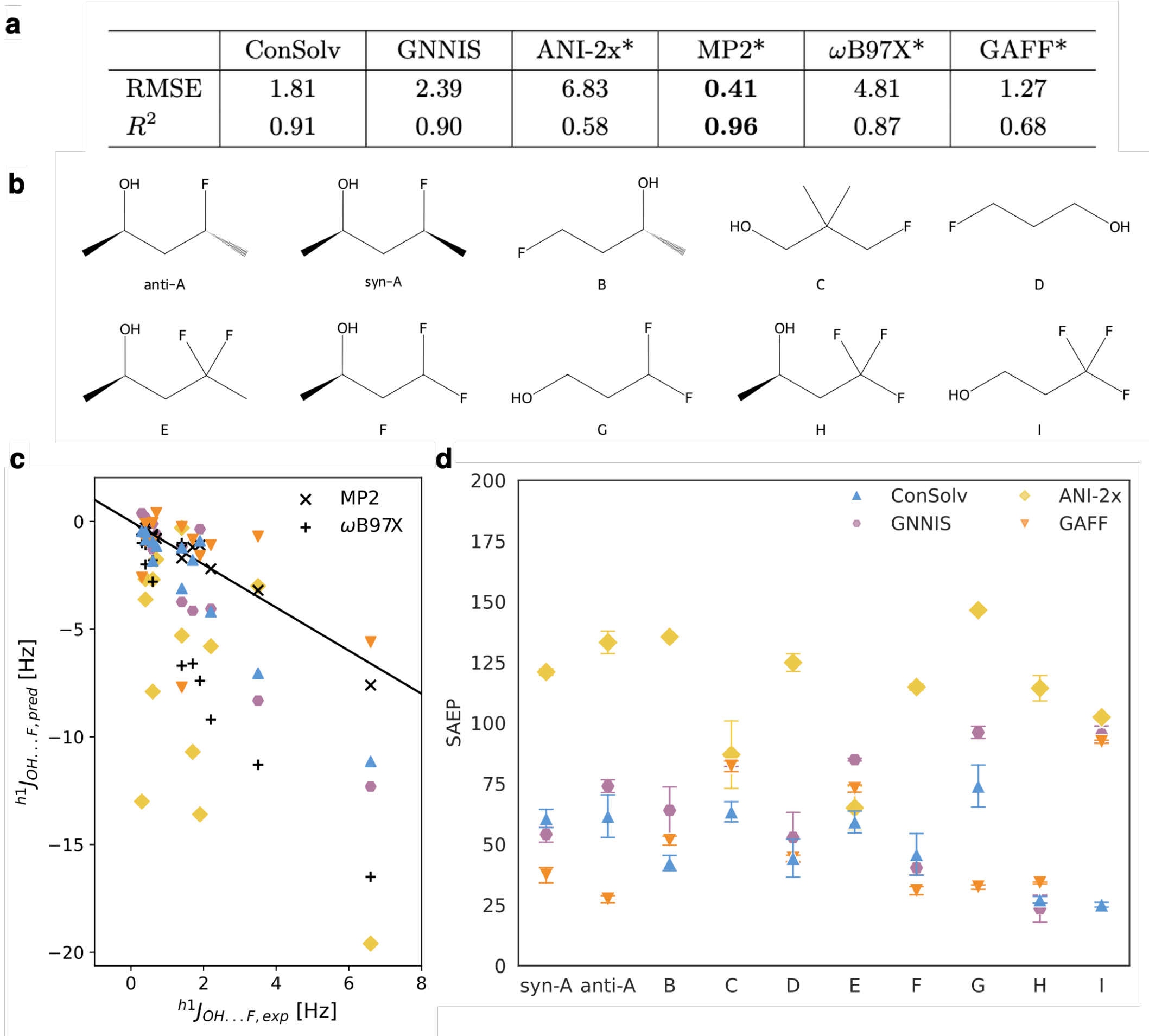}
    \caption{ Root mean squared error (RMSE), Pearson correlation coefficient $R^2$ (a) and parity plot (c) of  $^{h1}J_{OH...F}$ for 10 fluorohydrins (b) in chloroform solution with respect to experimental nuclear magnetic resonance (NMR) data. The black solid line denotes perfect correlation. Sum of absolute error of the populations (SAEP) (d) between the populations predicted by molecular dynamics simulations using ConSolv (blue), GNNIS (violet), ANI-2x (yellow), and GAFF (orange) models, and the Boltzmann populations calculated at the second-order M{\o}ller-Plesset perturbation theory (MP2)/6-311++G(2d,p)/polarizable continuum model (PCM) level of theory, labeled as MP2. $\omega$B97X denotes calculations with the $\omega$B97X/6-31G*/PCM level of theory.}
    \label{fig:saep}
\end{figure}

After extensive solvation free energy generalizability testing, we proceed to the out-of-target property test. To this end, we reproduce the test of Morado~\textit{et al.}~\cite{morado2023does} exploiting experimental nuclear magnetic resonance (NMR) data for 10 $\gamma$-fluorohydrins (Fig.~\ref{fig:saep}b) solvated in chloroform.
Note that the $\gamma$-fluorohydrin molecules are absent from both the SPICE and the Solv@TUM training set, providing a stringent test of model transferability. To calculate the J-coupling constants ($^{h1}J_{OH...F}$), we perform 20 MD simulations for each of the 10 $\gamma$-fluorohydrins starting with 20 different {RDKit-generated} configurations from SMILES strings. Each MD simulation is 1~ns long, totaling 20~ns of data per $\gamma$-fluorohydrin solute. Other MD simulation details are the same as for the solvation free energy calculations. We use MD trajectories to determine the relative conformational populations and subsequently compute a weighted average of the theoretical J-coupling constants obtained at the $\omega$B97X/6-311++G(2d,p)/PCM level of theory, as further detailed in the Supporting Information. We follow the same simulation and analysis protocols for ConSolv and the ML-based Generalized Born implicit solvent GNNIS model~\cite{katzberger2025rapid}. We additionally compare with previously published results~\cite{morado2023does} for other traditional and ML-based models.

ConSolv shows excellent performance across overall RMSE and $R^2$ error metrics, outperforming the explicit solvent classical force field GAFF on $R^2$, the ANI-2x model using an MLP for the solute and a classical force field for the solvent, as well as GNNIS (Figure~\ref{fig:saep}a). In addition, the predictions are in significantly closer agreement with the experiment than the ab initio implicit solvent $\omega$B97X/6-311++G(2d,p)/PCM model. Parity plot reveals several outliers for all models and the best-performing MP2/6-311++G(2d,p)/PCM model (Figure~\ref{fig:saep}c). 
We further quantify the quality of conformational sampling of $\gamma$-fluorohydrin molecules by computing the sum of absolute error of the populations (SAEP). It measures the difference between the model-generated populations against analytical Boltzmann populations derived at the MP2/6-311++G(2d,p)/PCM level of theory, with lower values indicating better compliance with reference quantum mechanics calculations. For analysis details, see Ref.~\cite{morado2023does}. {ConSolv achieves the closest agreement (average SAEP of 50.35) than previous ML-based ANI-2x (114.44) and GNNIS (67.00) models, and comparable performance to GAFF (50.50) (Figure~\ref{fig:saep}d). }

\subsection*{Insights into solvent effects via explainable artificial intelligence}
\begin{figure}[h!]
    \centering
    \includegraphics[width=1.0\linewidth]{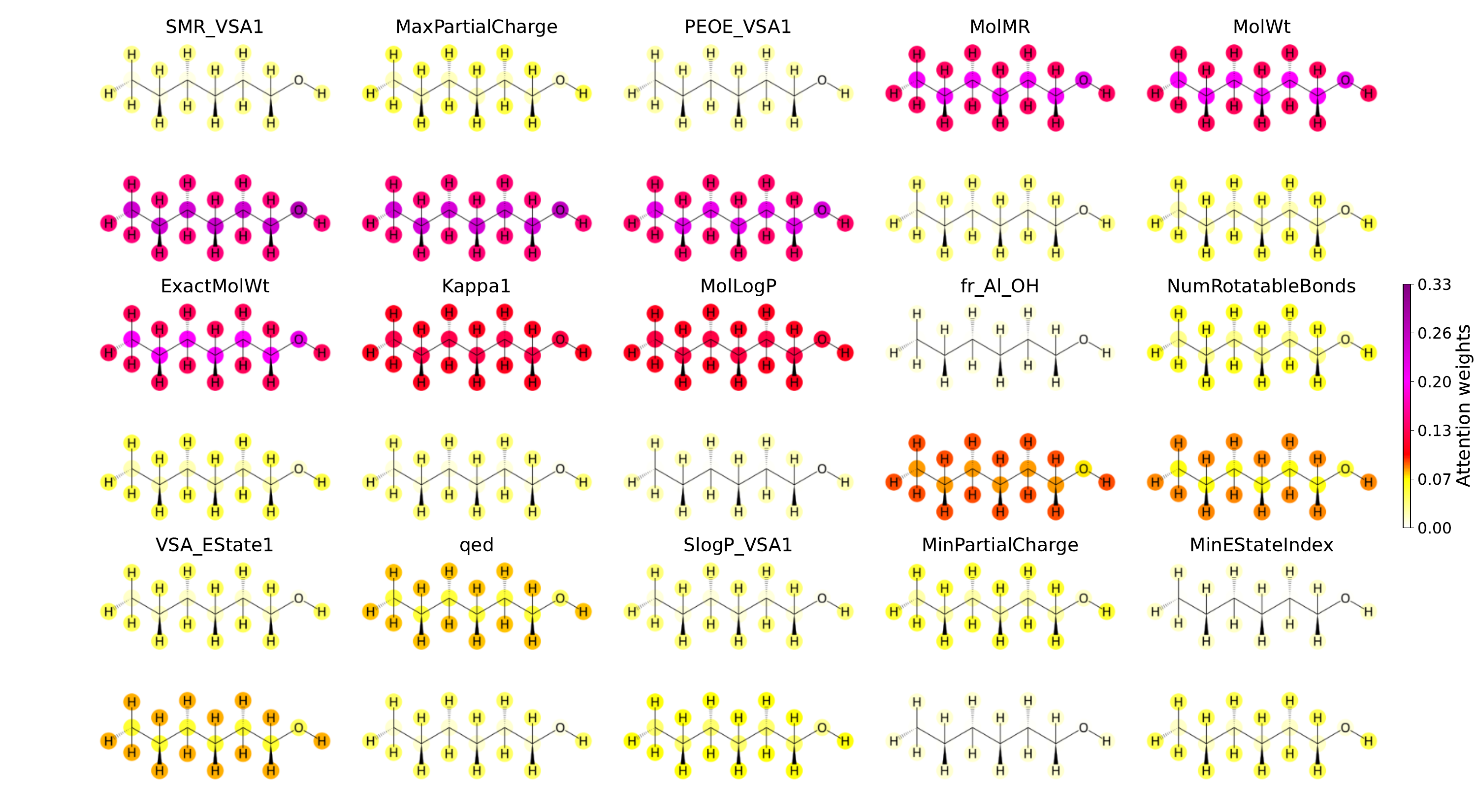}
    \caption{Per-atom attention weights (head$_3$) for 1-hexanol solute molecule and solvent descriptors of methanol (top, $\varepsilon=32.61$, polar) and 2,2,4-trimethylpentane (bottom, $\varepsilon=1.93$, non-polar).}
    \label{fig:attn_analysis}
\end{figure}

Owing to its attention-based design, ConSolv enables explainable artificial intelligence (AI) analysis. To understand how the solvent embedding block captures solvent effects, we visualize per-atom attention weights across four representative heads in the multi-head attention mechanism. {These weights reflect the learned affinity between solute atom-type embeddings and solvent descriptor representations, indicating which descriptors are preferentially selected during the aggregation of solvent information. Because the query entering this block is the initial atom-type embedding, the attention maps are independent of atomic coordinates and distinguish atom types rather than conformation-dependent local environments.} Note that the analysis does not directly quantify the modulation magnitude on individual node features, as this is jointly determined by the attention weights and the learned value representations.
Figure~\ref{fig:attn_analysis} shows the most interpretable head (head$_3$) for 1-hexanol in two solvents of contrasting polarity, namely methanol ($\varepsilon=32.61$, polar) with a predicted solvation free energy of $-6.2$~kcal/mol (experimental $-7.2$~kcal/mol) and 2,2,4-trimethylpentane (TMP, $\varepsilon=1.93$, non-polar) with a predicted solvation free energy of $-5.0$~kcal/mol (experimental $-5.1$~kcal/mol).

When predicting the potential energy of 1-hexanol in methanol, three size- and polarisability-related descriptors receive the highest attention weights, all sharply localized at the hydroxyl oxygen atom of the solute. These are the molecular refractivity (MolMR, which quantifies how easily the electron cloud of a
molecule is distorted by an external electric field and serves as a measure of molecular polarisability), the average molecular weight (MolWt), and the exact monoisotopic molecular weight (ExactMolWt). The attention weights for all three descriptors approach the observed maximum at the oxygen site, while carbon backbone atoms and carbon-bound hydrogen atoms show moderate attention weights. This pattern is physically consistent with the known behavior of polar protic solvation, where the refractivity and mass of the solvent govern the strength of inductive interactions and the energy required to form a solvent cavity~\cite{pierotti1976scaled}, and these contributions are most strongly felt at the electronegative, hydrogen-bond-accepting oxygen atom of 1-hexanol.

On the other hand, when TMP is used as a solvent, the attention pattern shifts markedly. The descriptors that receive the highest attention weights are still concentrated on hydroxyl oxygen, but they are now MaxPartialCharge (the largest positive partial charge on the solvent molecule) and PEOE\_VSA1 (the van der Waals surface area of the solvent in the most negatively charged partial charge bin, which reflects the distribution of electron density across the solvent surface). Both descriptors encode the near-absence of significant charge on the TMP molecule, and their localization at the hydroxyl oxygen of the solute reflects the electrostatic mismatch between the polar solute site and the charge-neutral solvent environment~\cite{reichardt2011solvents}. The attention thereby encodes the energetic penalty associated with placing a polar functional group in a medium with insufficient dielectric screening. 

A similar pattern as in head$_3$ is observed in head$_7$ ({Fig.~S7}), albeit with a smaller peak weight. {Head$_2$ likewise shows attention predominantly localized at the hydroxyl oxygen of 1-hexanol (Fig.~S8), although the dominant solvent descriptors differ between methanol and TMP. In methanol, MolMR (molar refractivity), MolWt, and ExactMolWt receive the highest attention at the oxygen site, indicating sensitivity to solvent size and polarizability-related properties. In TMP, the attention instead shifts toward SMR\_VSA1, MaxPartialCharge, and PEOE\_VSA1, which encode the distribution of polarizability and partial charge over the solvent surface. The remaining descriptors exhibit weaker and more broadly distributed attention across the carbon and hydrogen atoms.} The role of hydrogen-bond-specific descriptors is revealed in head$_5$ ({Fig.~S9}). In methanol, the descriptor fr\_Al\_OH (the count of aliphatic hydroxyl groups in the solvent molecule, which encodes its hydrogen-bond donor and acceptor capacity) achieves the highest attention weight in this head, with activation co-localized at the hydroxyl oxygen of 1-hexanol. This reflects the bifunctional nature of the hydroxyl group as simultaneously a hydrogen-bond donor and a hydrogen-bond acceptor, and is consistent with the dominance of hydrogen-bond complementarity in the solvation of alcohols by polar protic solvents~\cite{abraham1994hydrogen}. 
MinEStateIndex (the minimum electrotopological state index, which combines the electronic environment and topological connectivity of the least electron-rich atom in the molecule) shows additional moderate activation at the oxygen site in methanol, reinforcing the interpretation that electron-rich, topologically peripheral atoms constitute the primary sites of favorable electrostatic engagement with the polar solvent.
In TMP, head$_5$ instead localizes high attention for MinPartialCharge (the most negative partial atomic charge on any atom of the solvent molecule) exclusively at the hydroxyl oxygen of the solute, encoding the same electrostatic mismatch signal identified in head$_3$ from a complementary descriptor perspective.

The above analysis focuses on which molecular descriptors are most strongly attended when ConSolv evaluates solute solvation in different solvent environments. It is instructive to compare these learned descriptor preferences with the solvent properties used to parameterize DFT implicit-solvent models such as SM5, including surface tension, index of refraction, hydrogen bonding acidity, and hydrogen bonding basicity~\cite{li1999application}. Several connections emerge naturally. MolMR and SMR\_VSA1 are refractivity-based descriptors and therefore relate to the index of refraction and optical polarisability. Kappa1, MolWt, ExactMolWt, and van der Waals surface-area descriptors encode solvent size, shape, and exposed surface, which are indirectly related to cavity formation and nonpolar solvation terms that are often parameterized through surface-area or surface-tension contributions~\cite{cramer1999implicit}. The localized activation of fr\_Al\_OH provides the clearest molecular proxy for hydrogen bonding acidity and hydrogen bonding basicity in alcohol solvents, while MinEStateIndex, MaxPartialCharge, MinPartialCharge, and PEOE\_VSA1 provide complementary information on the local charge distribution that supports polar and hydrogen-bonding interactions~\cite{abraham1994hydrogen}. Thus, although ConSolv is not trained directly on these macroscopic solvent properties, the learned attention patterns recover several relevant physical categories through the available molecular descriptors.

Descriptors including the Quantitative Estimate of Drug-likeness (qed, a composite score developed to assess pharmaceutical candidates), the E-State-weighted van der Waals surface area (VSA\_EState1), and the logP-binned surface area (SlogP\_VSA1) remain consistently near or below the uniform baseline across all conditions, confirming their limited relevance to solvation free energy prediction.

\subsection*{Computational cost}

Finally, we compare the computational cost of ConSolv with the corresponding vacuum model without the solvent embedding block~\cite{batatia2022mace}, the published GNNIS implementation~\cite{katzberger2025rapid}, {and three explicit-solvent simulations using the same solvated configuration: OPLS-AA~\cite{jorgensen1996oplsaa}, GAFF~\cite{wang2004gaff}, and ANI-2x~\cite{devereux2020}.} The benchmark was carried out for the representative fluorohydrin anti-A molecule solvated in chloroform. {For all three explicit-solvent simulations, the same 4~nm simulation-box configuration containing 2418 atoms was used. For ANI-2x, the standard eight-member ensemble was used, with energies and forces averaged across all members; for GAFF, version~1.81 was used with AM1-BCC atomic partial charges~\cite{jakalian2002am1bcc} and particle-mesh Ewald (PME) treatment of long-range electrostatics~\cite{essmann1995pme}.} All timings were measured over 1~ns on an NVIDIA RTX~3090 GPU using a 1~fs timestep and exclude the initial compilation step. ConSolv achieved a simulation speed of 86~ns/day, compared with 92~ns/day for the vacuum model and 17~ns/day for GNNIS. {The corresponding explicit-solvent throughputs were 1310~ns/day for OPLS-AA, 2990~ns/day for GAFF, and 0.94~ns/day for ANI-2x.} These results demonstrate the efficiency of the current ConSolv implementation. Adding the solvent embedding block reduces the throughput by only about 7\% relative to the vacuum MACE model, indicating that solvent conditioning introduces only a small computational overhead. ConSolv is also about five times faster than GNNIS under the same simulation conditions. {Both ConSolv and GNNIS use GNN architectures with inexpensive solvent conditioning. However, GNNIS additionally requires geometry-dependent Born-radius calculations and generalized Born energy evaluation, adding overhead absent from ConSolv. Their similar interaction cutoffs (5 and 6 Å) and ConSolv’s larger parameter count (~0.6 vs.~0.1 million) suggest that neither cutoff nor model size explains ConSolv’s higher throughput. Instead, implementation differences, particularly JAX versus PyTorch and associated operation fusion and GPU kernel scheduling, may contribute substantially. Thus, the approximately five-fold speedup should be interpreted as an end-to-end implementation difference rather than an intrinsic architectural advantage or hardware-independent scaling law.} {Explicit-solvent simulations with ANI-2x are slower than classical force-field simulations because neural-network energy and force evaluations are more computationally demanding than prescribed analytical force-field expressions. In contrast, ConSolv substantially reduces the cost of ML-based simulations by eliminating explicit solvent molecules, making it approximately 90-fold faster than explicit-solvent ANI-2x. Although ConSolv remains more computationally expensive than classical force fields, its machine-learning representation offers the potential for higher accuracy while retaining much greater efficiency than explicit-solvent ML simulations.}

\section*{Conclusion}\label{sec:conclusion}

We have introduced ConSolv, a solvent-conditional MLP architecture for implicit solvent modeling. By incorporating solvent information into atomic embeddings via descriptor-based solvent conditioning, ConSolv enables accurate, computationally efficient predictions across chemically diverse solvent environments. Using a two-step training strategy, the model not only achieves strong correlation with experimental solvation free energies but also maintains simulation efficiency comparable to that of vacuum models. Our findings demonstrate that descriptor-conditioned solvent embeddings robustly capture solvent-specific effects, enabling reliable generalization to both polar and non-polar solvents, even those absent from the training set. In direct comparisons, ConSolv outperforms or matches traditional explicit-solvent force-field simulations
and DFT-based SCCS implicit-solvent approaches, all while operating at a fraction of the computational
cost of DFT-based implicit solvent methods. Furthermore, analysis of the solvent attention mechanism re-
veals chemically interpretable patterns: descriptors tied to solvent size and polarizability contribute broadly
to molecular representations, while electrostatic and hydrogen-bonding descriptors are selectively activated
around polar functional groups in a solvent-dependent fashion. Collectively, these results show that ConSolv
learns physically meaningful solvent–solute interaction patterns, underscoring its versatility and accuracy.

Despite ConSolv’s strong accuracy and transferability across diverse solvent environments, certain limitations persist. {The present model is designed for small, neutral organic and drug-like solutes represented by the filtered SPICE and Solv@TUM training sets~\cite{eastman2023spice,Hille2019}, together with pure non-aqueous solvents covered by the training descriptor distribution. Extending this framework to larger or highly flexible solutes, charged and ionic species, solvent mixtures, electrolytes, and other out-of-distribution systems represents a natural direction for future development.} The model’s predictive performance declines for solute chemistries that are underrepresented in the DFT training distribution, such as small gas-phase molecules, demonstrating that transferability is fundamentally linked to the chemical diversity of training data. Extending the model’s applicability to larger or more complex molecules will require access to broader solvation free energy datasets, such as CombiSolv~\cite{vermeire2021transfer}, or extensive force-labeled datasets for macromolecules in explicit solvent~\cite{eckwert2026transferable}. {The current model provides a foundation for future extensions to charged and ionic systems by incorporating total solute charge, long-range electrostatics, and broader intramolecular and solvation datasets~\cite{fuchs2025learning,sanocki2026generalization}. Extending ConSolv to solvent mixtures represents another promising direction, with composition-dependent training offering a potential route to capture non-ideal mixture effects beyond simple descriptor mixing. These developments will improve the model’s transferability to broader solute and solvent chemical spaces.} Moreover, further optimization of the classical solvent descriptor set could enhance both model performance and generalization. Thus, investigating alternative solvent representations, including pre-trained molecular embeddings from machine-learning-based feature-extraction models~\cite{oestreich2024small}, could potentially further enhance the model accuracy.

Importantly, the solvent-embedding architectural extension introduced in ConSolv is not restricted to the MACE architecture or the specific training framework utilized here. This approach can be readily incorporated into other MLPs and training protocols, broadening its potential impact. Particularly notable are emerging models that incorporate long-range interactions~\cite{fuchs2025learning, sanocki2026generalization}. Altogether, this work underscores the promise of solvent-conditioned MLPs for efficient, transferable implicit solvent modeling, providing a practical framework to advance molecular simulations across areas such as drug discovery, materials science, and electrochemistry.

\section*{Supporting Information}
Details of method implementation, training procedures, hyperparameters, and additional qualitative results are available in the Supporting Information.

\section*{Acknowledgements}
Funded by the European Union. Views and opinions expressed are however those of the
author(s) only and do not necessarily reflect those of the European Union or the European Research Council
Executive Agency. Neither the European Union nor the granting authority can be held responsible for them. This
work was funded by the ERC (StG SupraModel) - 101077842. Funded by the Deutsche Forschungsgemeinschaft (DFG, German Research Foundation) – SPP2363, Projektnummer: 561190767. The authors gratefully acknowledge the Gauss Centre for Supercomputing e.V. (www.gauss-centre.eu) for funding this project by providing computing time through the John von Neumann Institute for
Computing (NIC) on the GCS Supercomputer JUPITER and JUWELS~\cite{JUWELS} at Jülich Supercomputing Centre (JSC).

\section*{Data and Software Availability}
The source code used in this study is publicly available on
GitHub at  \url{https://github.com/tummfm/ConSolv}. The model files and data supporting this study are publicly available on Zenodo at  \url{https://zenodo.org/records/22956180}.

\printbibliography

\clearpage
\begin{center}
  {\bfseries TOC Graphic\par}
  \vspace{8pt}
  \includegraphics[width=3.25in,height=1.75in,keepaspectratio]{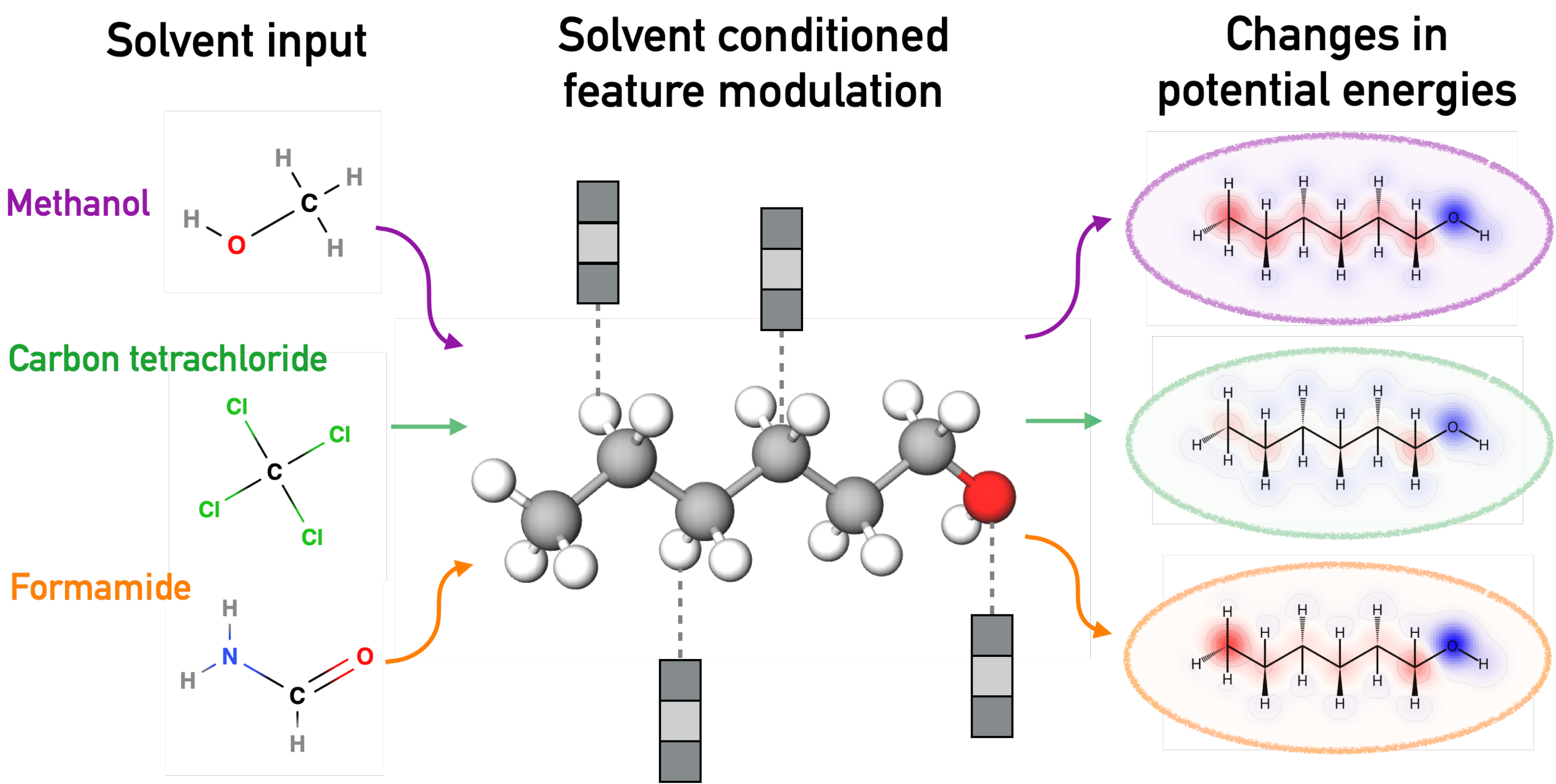}
\end{center}

\end{document}